\documentclass[fleqn,12pt,twoside]{article}
\usepackage[headings]{espcrc1}
\usepackage{epsfig}

\readRCS
$Id: espcrc1.tex,v 1.2 2004/02/24 11:22:11 spepping Exp $
\newcommand{\AmS}{{\protect\the\textfont2
  A\kern-.1667em\lower.5ex\hbox{M}\kern-.125emS}}

\title{System size, energy and pseudorapidity dependence of directed and
elliptic flow at RHIC}

\author{
S.~Manly\thanks{Presented at the 18$^{th}$ International
Conference on Ultra-Relativistic Nucleus-Nucleus Collisions, in
Budapest, Hungary, August 4-9, 2005.} for the PHOBOS
Collaboration:
\\
B.Alver$^4$, B.B.Back$^1$, M.D.Baker$^2$, M.Ballintijn$^4$,
D.S.Barton$^2$, R.R.Betts$^6$, A.A.Bickley$^7$, R.Bindel$^7$,
A.Budzanowski$^3$, W.Busza$^4$, A.Carroll$^2$, Z.Chai$^2$,
V.Chetluru$^6$, M.P.Decowski$^4$, E.Garc\'{\i}a$^6$, T.Gburek$^3$,
N.George$^2$, K.Gulbrandsen$^4$, S.Gushue$^2$, C.Halliwell$^6$,
J.Hamblen$^8$, G.A.Heintzelman$^2$, C.Henderson$^4$,
I.Harnarine$^6$, D.J.Hofman$^6$, R.S.Hollis$^6$, R.Ho\l
y\'{n}ski$^3$, B.Holzman$^2$, A.Iordanova$^6$, E.Johnson$^8$,
J.L.Kane$^4$, N.Khan$^8$, W.Kucewicz$^6$, P.Kulinich$^4$,
C.M.Kuo$^5$, W.Li$^4$, W.T.Lin$^5$, C.Loizides$^4$, S.Manly$^8$,
A.C.Mignerey$^7$, R.Nouicer$^{2,6}$, A.Olszewski$^3$, R.Pak$^2$,
I.C.Park$^8$, C.Reed$^4$, L.P.Remsberg$^2$, M.Reuter$^6$,
E.Richardson$^7$, C.Roland$^4$, G.Roland$^4$, L.Rosenberg$^4$,
J.Sagerer$^6$, P.Sarin$^4$, P.Sawicki$^3$, I.Sedykh$^2$,
W.Skulski$^8$, C.E.Smith$^6$, M.A.Stankiewicz$^2$,
P.Steinberg$^2$, G.S.F.Stephans$^4$, A.Sukhanov$^2$,
A.Szostak$^2$, J.-L.Tang$^5$, M.B.Tonjes$^7$, A.Trzupek$^3$,
C.Vale$^4$, G.J.van~Nieuwenhuizen$^4$, S.S.Vaurynovich$^4$,
R.Verdier$^4$, G.I.Veres$^4$, P.Walters$^8$,E.Wenger$^4$,
D.Willhelm$^2$, F.L.H.Wolfs$^8$, B.Wosiek$^3$, K.Wo\'{z}niak$^3$,
A.H.Wuosmaa$^1$, S.Wyngaardt$^2$,
B.Wys\l ouch$^4$\\
$^1$~Argonne National Laboratory, Argonne, IL 60439-4843, USA\\
$^2$~Brookhaven National Laboratory, Upton, NY 11973-5000, USA\\
$^3$~Institute of Nuclear Physics PAN, Krak\'{o}w, Poland\\
$^4$~Massachusetts Institute of Technology, Cambridge, MA
02139-4307,
USA\\
$^5$~National Central University, Chung-Li, Taiwan\\
$^6$~University of Illinois at Chicago, Chicago, IL 60607-7059, USA\\
$^7$~University of Maryland, College Park, MD 20742, USA\\
$^8$~University of Rochester, Rochester, NY 14627, USA\\
}

\runtitle{System size, energy and $\eta$ dependence of directed
and elliptic flow at RHIC} \runauthor{S. Manly}

\begin{document}

\maketitle

PHOBOS measurements of elliptic flow are presented as a function
of pseudorapidity, centrality, transverse momentum, energy and
nuclear species.  The elliptic flow in Cu-Cu is surprisingly
large, particularly for the most central events. After scaling out
the geometry through the use of an alternative form of
eccentricity, called the participant eccentricity, which accounts
for nucleon position fluctuations in the colliding nuclei, the
relative magnitude of the elliptic flow in the Cu-Cu system is
qualitatively similar to that measured in the Au-Au system.

\section{Introduction}
The characterization of the collective flow of produced particles
by their azimuthal anisotropy has proven to be one of the more
fruitful probes of the dynamics of heavy ion collisions at RHIC.
In particular, differential flow measurements provide information
crucial in constraining three-dimensional hydrodynamic models of
relativistic heavy ion collisions.

This work presents new results on data taken by the PHOBOS
experiment at RHIC showing a detailed comparison of differential
measurements of flow across species. The data were taken with the
PHOBOS experiment at RHIC during Au-Au and Cu-Cu collisions
ranging over an order of magnitude in collision energy. The PHOBOS
detector employs silicon pad detectors to perform tracking, vertex
detection and multiplicity measurements. Details of the setup and
the layout of the silicon sensors can be found
elsewhere~\cite{phobos_det}. The Au-Au data shown here are from
previous work~\cite{limfrag}\cite{ptpaper}. The presented Cu-Cu
data are analyzed in a similar fashion.

\section{Results }


PHOBOS has recently completed measurements of the directed flow
signal, v$_{1}$, as a function of pseudorapidity ($\eta$) in Au-Au
collisions at $\sqrt{s_{_{NN}}} =$200, 130, 62.4 and 19.6 GeV.
These measurements were performed with the standard subevent
technique described in reference~\cite{qm04} and confirmed with a
mixed harmonic method~\cite{starmixedharm}.  The directed flow
signal exhibits extended longitudinal scaling in a fashion
analogous to that seen in the elliptic flow~\cite{limfrag}.  These
results are presented in reference~\cite{Aliceposterpaper}.

\begin{figure}
\begin{minipage}{18pc}
\includegraphics[width=18pc]{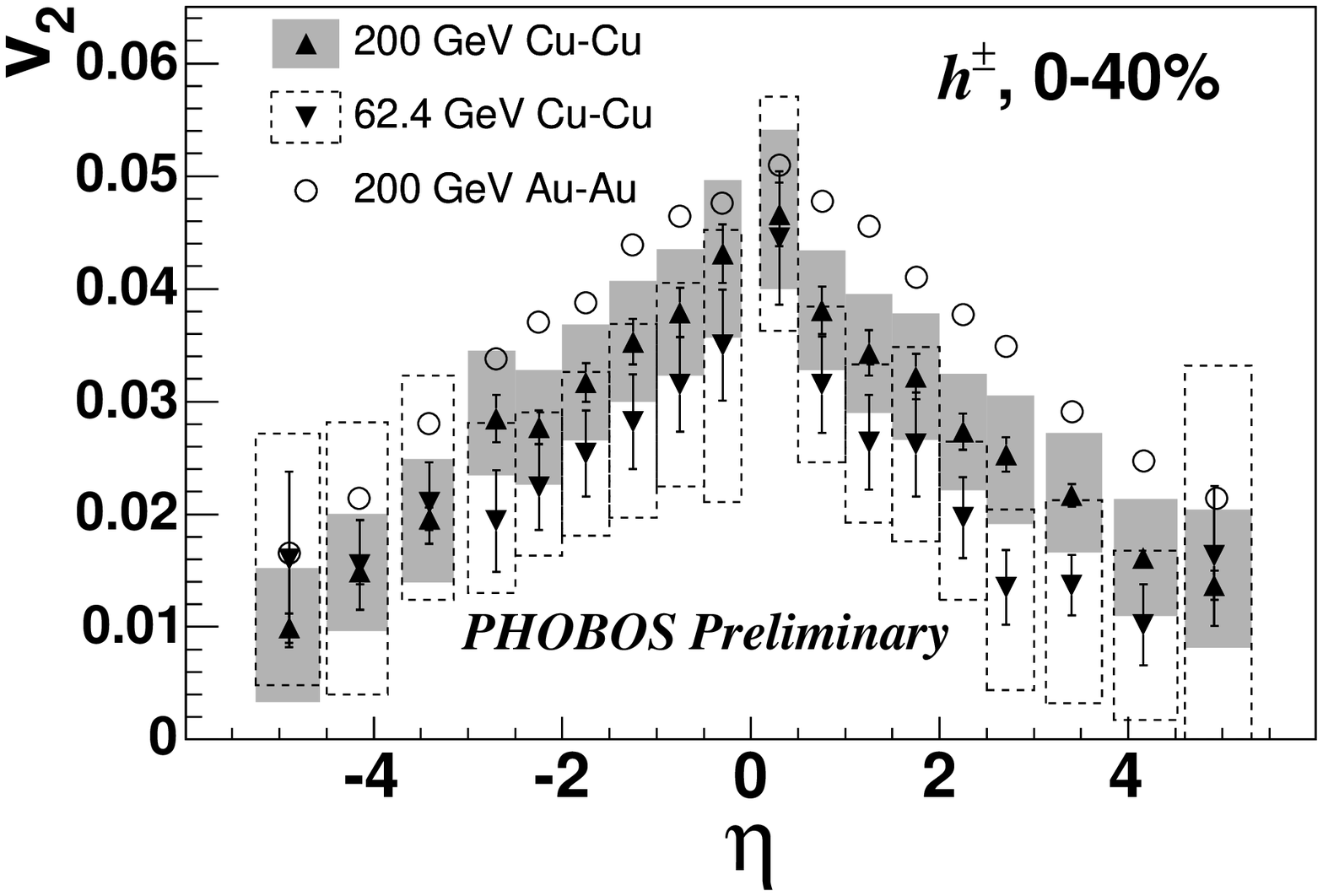}
 \caption{v$_{2}$ vs.\ $\eta$ for Cu-Cu collisions at
 $\sqrt{s_{_{NN}}} =$200 and 62.4 GeV.  The boxes
 show the 90\% C.L. systematic errors and the
 bars represent the 1-$\sigma$ statistical errors.
 Previously published Au-Au data (without error bars)
 is shown for comparison.}
\label{v2etacu}
\end{minipage}
\hspace{2pc}
\begin{minipage}{18pc}
\vspace{-0.5in}
\includegraphics[width=18pc]{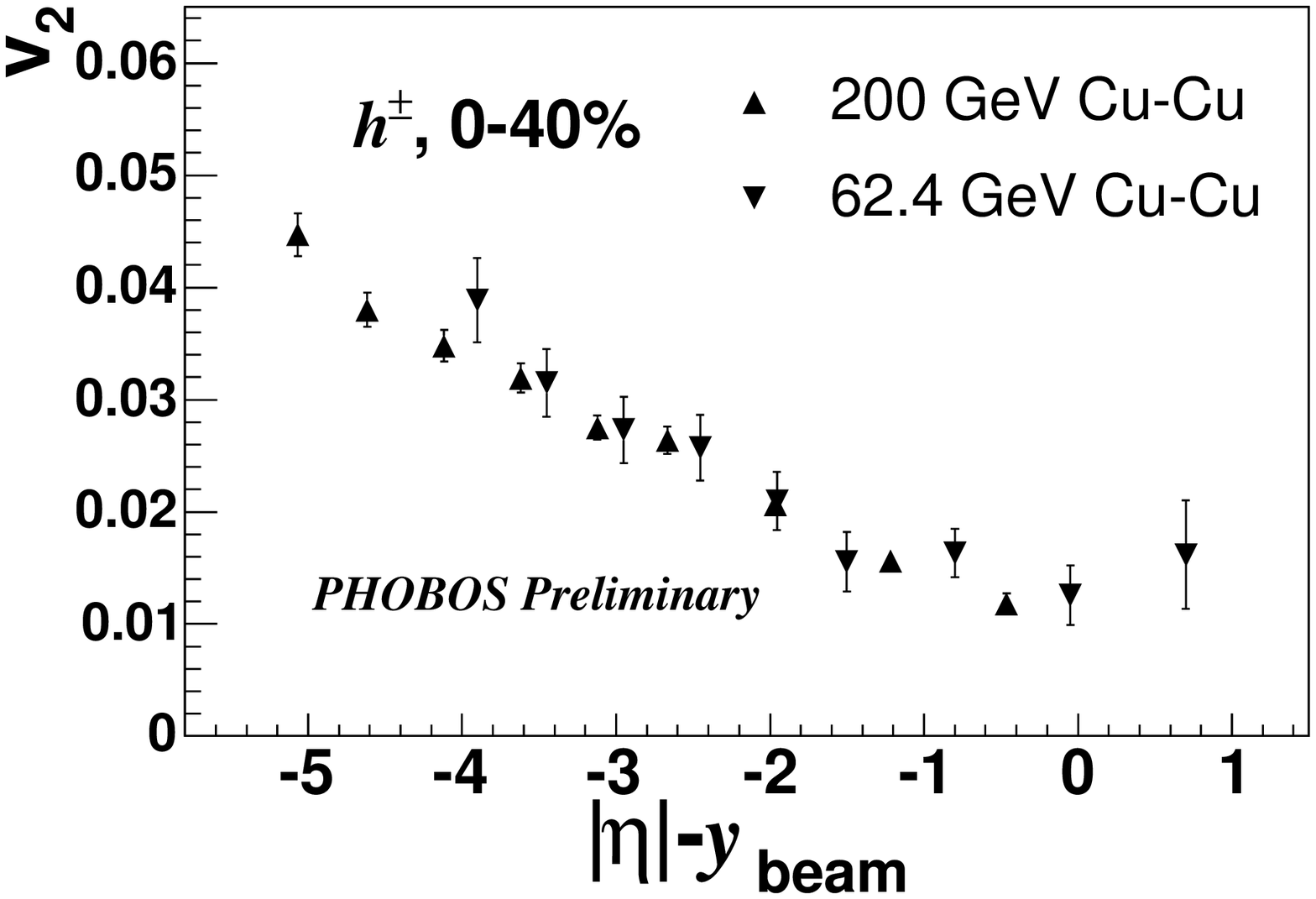}
 \caption{v$_{2}$ vs.\ $|\eta|-$y$_{beam}$ for Cu-Cu collisions at
 $\sqrt{s_{_{NN}}} =$200 and 62.4 GeV.
 Only statistical errors are shown.} \label{v2etapcu}
\end{minipage}
\end{figure}
\begin{figure}
\begin{minipage}{18pc}
\includegraphics[width=18pc]{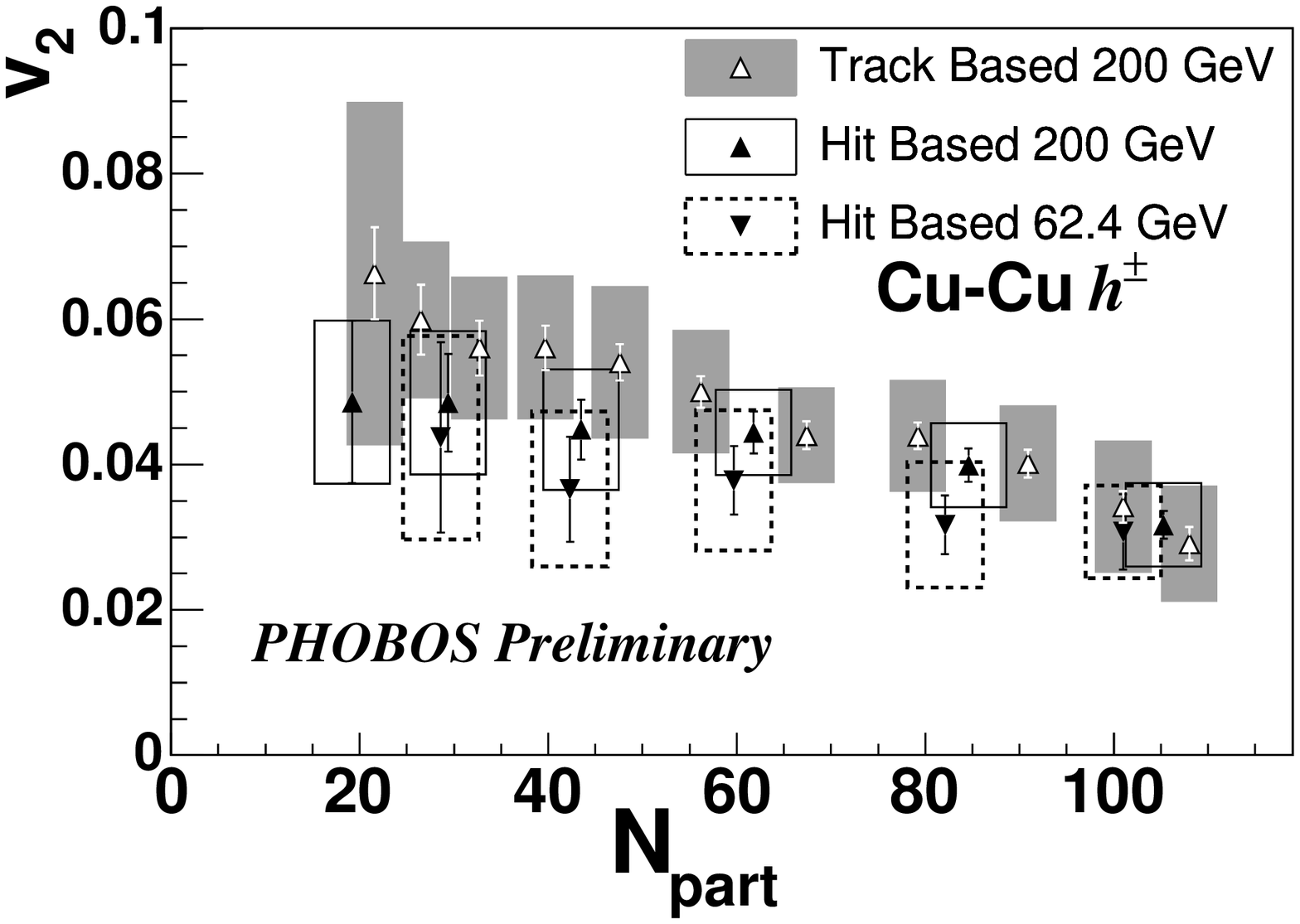}
 \caption{v$_{2}$ vs.\ N$_{\rm part}$ for Cu-Cu collisions at
 $\sqrt{s_{_{NN}}} =$200 and 62.4 GeV.  The boxes
 show the 90\% C.L. systematic errors and the
 lines represent the 1-$\sigma$ statistical errors.
 The results from two analysis methods are
 shown, similar to those presented in~\cite{limfrag} and~\cite{ptpaper}.}
\label{v2npartcu}
\end{minipage}
\hspace{2pc}
%
\begin{minipage}{18pc}
\includegraphics[width=18pc]{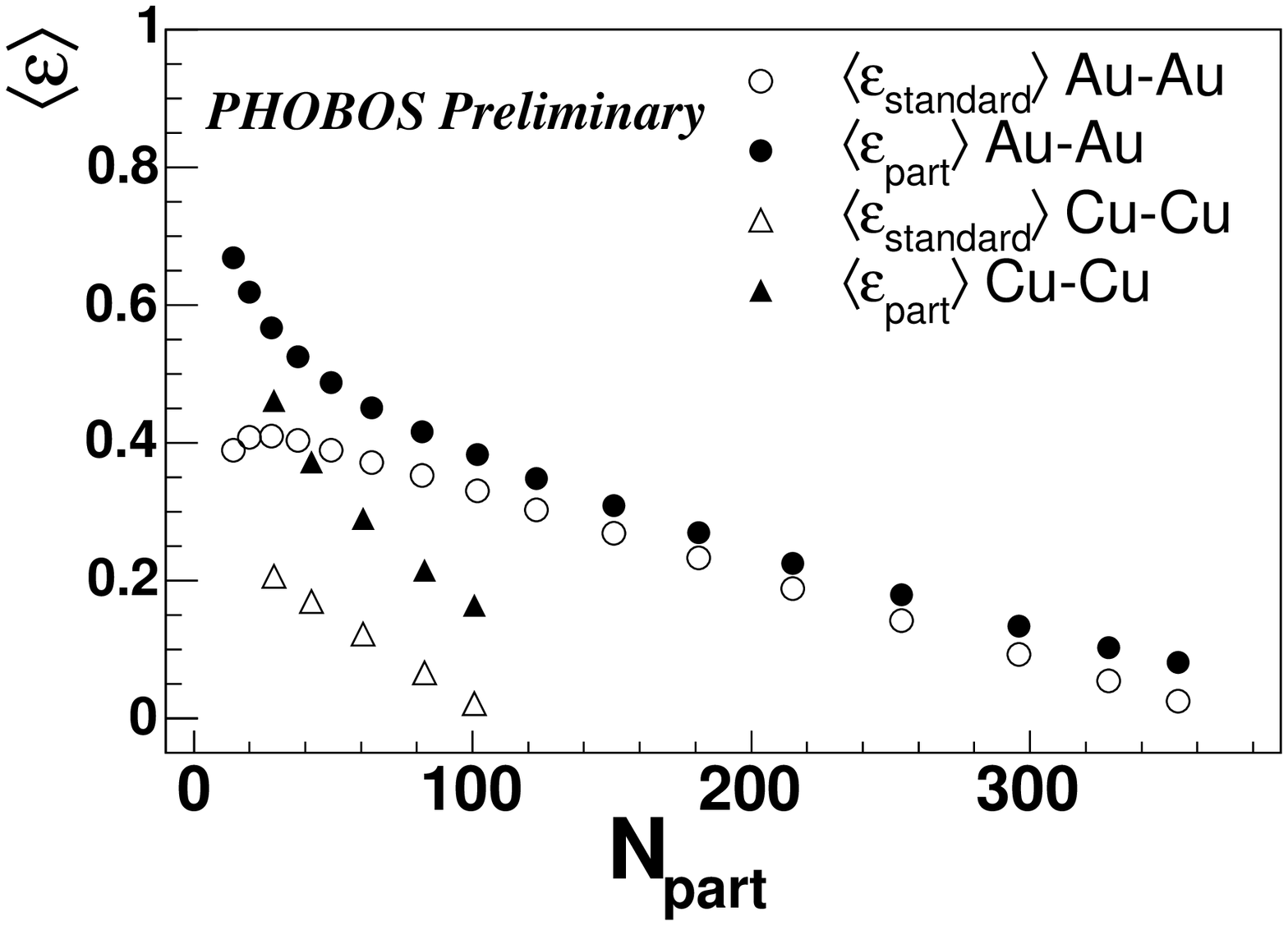}
 \caption{The average eccentricity defined in two ways
 ($\langle \varepsilon_{\rm standard} \rangle$ and
 $\langle \varepsilon_{\rm part} \rangle$), as described
 in the text, vs.\ N$_{\rm part}$ for simulated Au-Au and Cu-Cu
 collisions at $\sqrt{s_{_{NN}}} =$200 GeV.
 Only statistical errors are shown.} \label{aveecc}
\end{minipage}
\end{figure}

Figure~\ref{v2etacu} shows the elliptic flow signal, v$_{2}$, as a
function of $\eta$ in Cu-Cu collisions at $\sqrt{s_{_{NN}}} =$200
and 62.4 GeV for the 40\% most central collisions.  The
resemblance to published Au-Au results (also shown in
Figure~\ref{v2etacu}) is striking~\cite{limfrag}. The Cu-Cu system
also exhibits extended longitudinal scaling, as shown in
Figure~\ref{v2etapcu}, and as already seen in Au-Au
collisions~\cite{limfrag}. The centrality dependence of v$_{2}$ is
presented in Figure~\ref{v2npartcu}. Substantial flow is present
in the Cu-Cu signal for even the most central events.

\begin{figure}[htb]
\vspace{-0.2in} \epsfxsize=1.0\textwidth\epsfbox{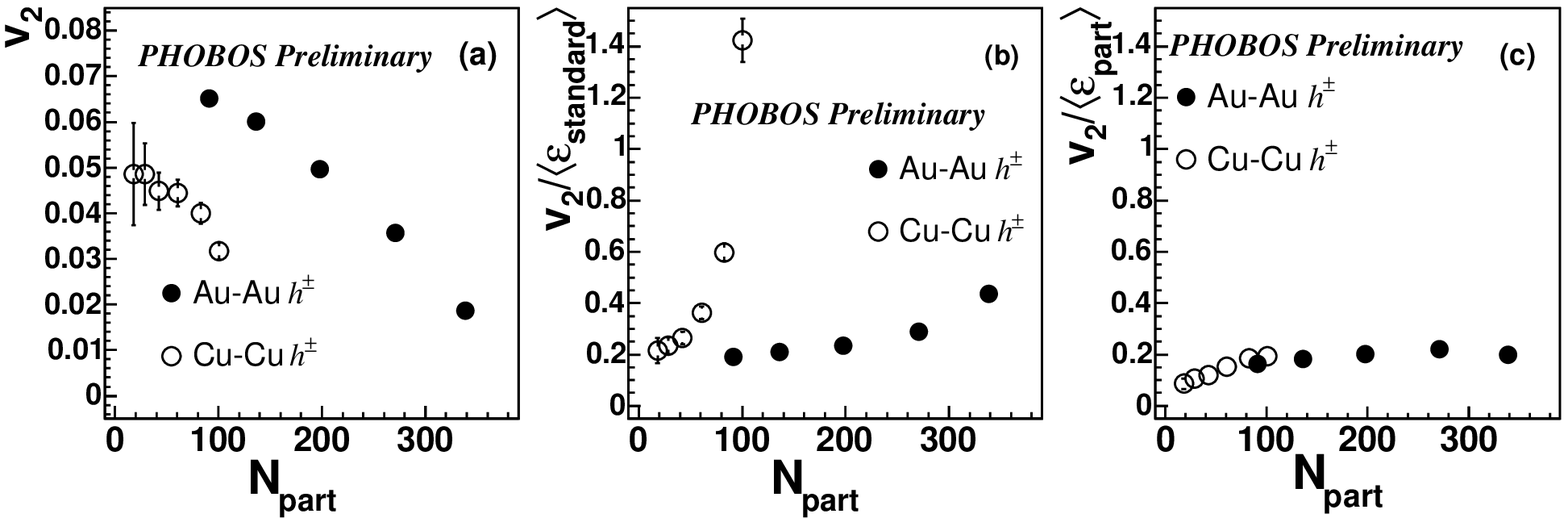}
 \caption{(a)  v$_{2}$ (unscaled) vs.\ N$_{\rm part}$,
 (b) v$_{2}$/$\langle \varepsilon_{\rm standard} \rangle$
 vs.\ N$_{\rm part}$ and (c) v$_{2}$/$\langle \varepsilon_{\rm part}
 \rangle$
 vs.\
 N$_{\rm part}$,
 for
 Cu-Cu and Au-Au collisions at $\sqrt{s_{_{NN}}} =$200 GeV.
 Only statistical errors are shown.}
\label{v2ecc2figs}
\end{figure}

\begin{figure}[htb]
\vspace{-0.2in}
 \epsfxsize=1.0\textwidth\epsfbox{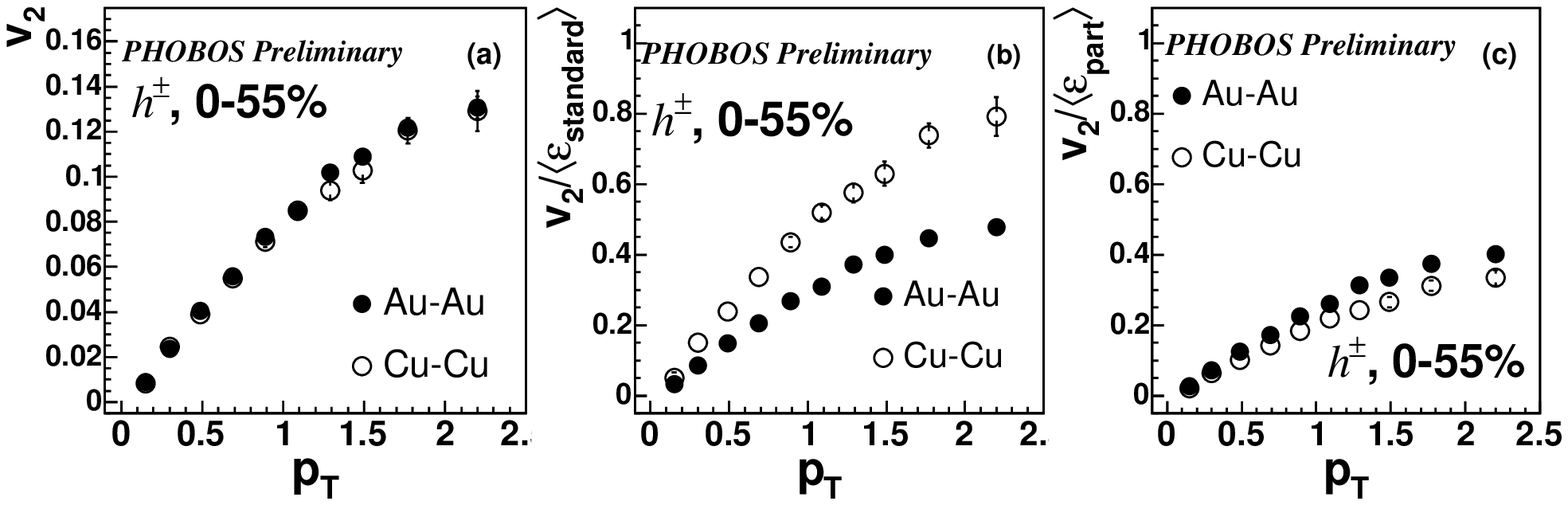}
 \caption{ (a) v$_{2}$ (unscaled) vs.\ p$_{T}$, (b)
 v$_{2}$/$\langle \varepsilon_{\rm standard} \rangle$ vs.\ p$_{T}$ and (c)
v$_{2}$/$\langle \varepsilon_{\rm part} \rangle$ vs.\ p$_{T}$,
 for
 the 55\% most central Cu-Cu and Au-Au collisions at
 $\sqrt{s_{_{NN}}} =$200 GeV.
 Only statistical errors are shown.}
\label{v2_ptstd}
\end{figure}


In order to compare flow signals across nuclear species it is
important to scale out the difference in the initial geometric
asymmetry of the collision, i.e., the eccentricity of the
collision. This is crucial since for a selected centrality range,
the average eccentricity depends on the size of the colliding
species.  Typically, the eccentricity is defined by relating the
impact parameter of the collision in a Glauber model simulation to
the eccentricity calculated assuming the minor axis of the overlap
ellipse to be along the impact parameter vector. Thus, if the
$x$-axis is defined to be along the impact parameter vector and
the $y$-axis perpendicular to that in the transverse plane, the
eccentricity is determined by
$ \varepsilon=\frac{\sigma_{y}^{2}-\sigma_{x}^{2}}
 {\sigma_{y}^{2}+\sigma_{x}^{2}}$,
where $\sigma_{x}$ and $\sigma_{y}$ are the RMS widths of the
participant nucleon distributions projected on the $x$ and $y$
axes, respectively. Let us call the eccentricity determined in
this fashion $\varepsilon_{\rm standard}$.

For small systems or small transverse overlap regions,
fluctuations in the nucleon positions frequently create a
situation where the minor axis of the ellipse in the transverse
plane formed by the participating nucleons is not along the impact
parameter vector. One way to address this issue is to make a
principal axis transformation, rotating the $x$ and $y$ axes used
in the eccentricity definition in the transverse plane in such a
way that $\sigma_{x}$ is minimized. Let us call the eccentricity
determined in this fashion $\varepsilon_{\rm part}$. In terms of
the original $x$ and $y$ axes (in fact, any pair of perpendicular
transverse axes),
\begin{equation}
 \varepsilon_{\rm part}=\frac{
\sqrt{(\sigma_{y}^{2}-\sigma_{x}^{2})^{2}+4(\sigma_{xy})^{2}} }
 {\sigma_{y}^{2}+\sigma_{x}^{2}}.
\end{equation}
In this formula, $\sigma_{xy}=\langle xy \rangle - \langle x
\rangle \langle y \rangle$. The average values of
$\varepsilon_{\rm standard}$ and $\varepsilon_{\rm part}$ are
quite similar for all but the most peripheral interactions for
large species, as is shown in Figure~\ref{aveecc} for Au-Au.  For
smaller species such as Cu, however, fluctuations in the nucleon
positions become quite important for all centralities and the
average eccentricity can vary significantly depending on how it is
calculated.  This is also illustrated in Figure~\ref{aveecc}. It
is worth noting that the Glauber model used in these calculations
does not include an excluded volume for the nucleons.

\section{Summary}

The crucial importance of the definition of eccentricity in
comparing Au-Au and Cu-Cu results can be seen in
Figures~\ref{v2ecc2figs}-\ref{v2_ptstd}, where various comparisons
are made between Au-Au and Cu-Cu data using the
eccentricity-scaled elliptic flow. Given the qualitative and
quantitative similarities between the results in the Au-Au and
Cu-Cu systems when scaled by $\langle \varepsilon_{\rm part}
\rangle$, it seems likely that $\varepsilon_{\rm part}$ or a
rather similar quantity is the relevant eccentricity for the
azimuthal expansion. Perhaps more interesting is the fact that
these data show that qualitative features attributed to collective
effects in Au-Au persist down to the relatively small numbers of
participants seen in the Cu-Cu collision and are of comparable
magnitude.


\end{document}